# A sampling approach for the estimation of the critical parameters of the SARS-CoV-2 epidemic: an operational design

Giorgio Alleva[1]  Giuseppe Arbia [2]  Piero Demetrio Falorsi[3]  Vincenzo Nardelli[4]
Alberto Zuliani[5]

**Abstract:** Given the urgent informational needs connected with the diffusion of infection with regard to the COVID-19 pandemic, in this paper, we propose a sampling design for building a continuous-time surveillance system. Compared with other observational strategies, the proposed method has three important elements of strength and originality: (i) it aims to provide a snapshot of the phenomenon at a single moment in time, and it is designed to be a continuous survey that is repeated in several waves over time, taking different target variables during different stages of the development of the epidemic into account; (ii) the statistical optimality properties of the proposed estimators are formally derived and tested with a Monte Carlo experiment; and (iii) it is rapidly operational as this property is required by the emergency connected with the diffusion of the virus. The sampling design is thought to be designed with the diffusion of SAR-CoV-2 in Italy during the spring of 2020 in mind. However, it is very general, and we are confident that it can be easily extended to other geographical areas and to possible future epidemic outbreaks. Formal proofs and a Monte Carlo exercise highlight that the estimators are unbiased and have higher efficiency than the simple random sampling scheme.

**Some keywords:** COVID-19 diffusion; relative efficiency; epidemic monitoring; health surveillance system; sampling design; unbiasedness.

1. Background and purpose

The urgent, worldwide need for a method of controlling the spread of SARS-CoV-2 requires an accurate evaluation of the sources of data on which the estimation of the epidemic's main parameters can be based. Only in this way will we be able to monitor the evolution of the epidemic over time while simultaneously supporting decision makers in evaluating the effects of the restrictive measures gradually introduced to try and stop the spread and the time required for the reduction and removal of these measures. In general, this approach enables the production of future forecasts of the evolution of the disease, and these forecasts are the essential basis for achieving an effective healthcare response. Indeed, while some degree of uncertainty is inherent in any statistical model, the level of inaccuracy in terms of monitoring the development of the situation can and must be kept under control.

---

[1] Sapienza University of Rome, giorgio.alleva@uniroma1.it
[2] Catholic University of the Sacred Heart, Milan
[3] Former Director of Methodology at Istat and International Consultant
[4] Bicocca University, Milan
[5] Emeritus Professor, Sapienza University of Rome

The objective of the proposed method is the definition of an observational protocol for observing the SARS-CoV-2 epidemic over time and providing statistically unbiased and efficient estimates of the sizes of the different attributes of any population identified as a concern with regard to the epidemic. Moreover, we aim to propose a dynamic monitoring tool that can be suitably calibrated both in the growth phase of the infection rate and in the decline phase, with estimates extended to the parameters of the progressive immunization model for the population. All estimates can be produced with associated reliability measures.

However, apart from a few remarkable exceptions, until now, the data that have been collected favour the examination of cases in which the patients display symptoms. This situation is described in statistics as "convenience sampling", and no sound probabilistic inference is possible under such a sampling approach (Hansen et al., 1953). More precisely, in a formal sample design, the choices of observations are suggested by a precise mechanism based on the definition of the inclusion probabilities of each unit (and, hence, by a sound probabilistic inference method); in contrast, with a convenience sampling, no probabilities of inclusion can be calculated, thus giving rise to over- or underrepresentation of the sample units.

In particular, several studies on COVID-19 diffusion have clearly shown (e.g., Aguilar et al., 2020; Chugthai et al., 2020; Li et al., 2020; Mizumoto et al., 2020a, 2020b and Yelin et al., 2020) that the available data strongly underestimate the number of infected people that they are unable to capture, e.g., asymptomatic cases with an obvious overestimation of the lethality rate[6]. On the other hand, a broad data collection method using medical swabs that is carried out on a voluntary basis does not constitute a probabilistic sample either[7]. For instance, the practice of systematically collecting observations from people in the vicinity of supermarkets leads to an overinclusion of healthy people in the sample and to a systematic exclusion of those who (either because they are manifesting symptoms or because they feel weak) have chosen to stay confined at home.

However, it is of crucial importance for government and health officials and for the general population to have a clear understanding of the dynamics of an epidemic while it is in progress so that the government can take appropriate measures and guide individual behaviours. In such a situation, it is essential to set up a data collection system that can provide unbiased estimates and statistically valid comparisons over time and across different geographic areas.

For sampling during an epidemic to be empirically relevant, any data collection design must be technically specified, the properties of the associated estimators have to be proved formally, and the design has also to satisfy the following two conditions:

---

[6] The lethality rate is given by the number of deaths out of the total number of infected people.
[7] https://www.theguardian.com/world/2020/mar/30/immunity-passports-could-speed-up-return-to-work-after-covid-19.

- It has to be implemented as a surveillance system (or strictly related to an existing one) and repeated in several waves rather than as a one-time survey.
- It has to be immediately operational considering the practical implications of the collected data.

The latter point is particularly relevant to the idea that the task may prove challenging, especially in a situation where all health operators are employed full time in emergency operations related to the care of the most severely infected people.

Rather surprisingly, the literature on this subject is still extremely poor. Few contributions have suggested the use of crowdsourced data rather than a sampling design along with officially collected data (Leung and Leung, 2020; Sun et al., 2020); the risk of erroneous inferences based on these data has been pointed out by Arbia (2020), Di Gennaro et al. (2020) and Ioannidis (2020). Our aim is to suggest a sampling design whose statistical optimality properties are formally proven, where the design is also operational and can be immediately put into action upon taking the many practical obstacles that may arise in an emergency into account. Although we have the Italian COVID-19 situation in mind, we are rather confident that the suggested protocol could be easily extended to other countries.

The rest of the paper is organized as follows. In Section 2, we present a review of the strategies and experiences in progress with regard to data collection until early April 2020. In Section 3, we present the basic sampling framework of our suggested design by distinguishing two subsets of the population to be surveyed, namely, those in which a state of infection has already been verified and those who were in contact with them (group A) and healthy persons (group B). The different roles of the two groups in monitoring infections during different stages of the epidemic are also discussed. In Section 4, we focus on the parameters of interest that we aim at measuring with the suggested sampling design based on the two groups, and we discuss how to disentangle possible overlaps between them whose presence may undermine the statistical properties of the estimations. In Section 5, we provide a general description of the sampling schemes for the two groups and the various operations to be realized. In Section 6, we prove the unbiasedness of the estimates and derive the expressions of the sampling variances. Section 7 is devoted to envisaging an extension of the proposed methodology to subsequent waves of data collection for the purpose of monitoring phenomena at different moments of time and during different stages of the epidemic. Section 8 contains a discussion on the efficiency of the estimators. Section 9 illustrates the empirical results of a simulation study. Finally, in Section 10, we suggest some practical implications of the study and future research priorities. All formal proofs can be found in the Appendix.

## 2. Data collection during an epidemic: a review of strategies and experiences currently in progress

In the emergency phase connected with the quick and uncontrolled diffusion of COVID-19, governments and institutions in charge are fully aware that knowledge and understanding of the dynamics at work represent the central element for establishing how to intervene and in which geographical areas intervention is most urgent.

In reviewing the approaches followed by various countries until early April 2020, we can identify four strategies and experiences in progress with regard to the estimation of the disease phenomena in the entire population.

a) The first consists of *massive test campaigns* (regardless of the presence of symptoms) carried out without following a formal sampling design; these are essentially aimed at intervening during outbreaks of the epidemic to identify subjects who are infected but with no symptoms or only slight symptoms. This was the strategy of South Korea and Hong Kong, as well as of the United Arab Emirates, Australia, Iceland, and the Veneto Region in Italy[8]. The big limit of this approach is the impossibility of making statistical inferences for the whole population based on the results.

b) The second possible strategy consists of *diagnostic tests through a probabilistic sample* according to a planned design for the estimation of the phenomena of interest with predetermined precision levels. This approach is aimed at estimating the effective amount of infections, including those in the asymptomatic population. This approach was used in the project performed by the Helmholtz Center for Research on Infections in Germany; this project was based on testing patients' blood for antibodies to the Covid-19 pathogen and involved over 100,000 individuals (Hackenbroch, 2020). Similarly, in Romania, a random sample of 10,500 people living in Bucharest has been planned to detect infected persons by following the directions of the Matei Bals Institute of Infectious Diseases in Bucharest (Romania-insider.com, 2020). Finally, a random selection of people who do not meet the testing criteria will be observed at two Canberra locations by the Australian Capital Territory (Abc, 2020). All these sample surveys are cross-sectional and useful for measuring the infection rate at a precise instant. However, they have distinct characteristics from those of continuous panel-type surveys with rotated samples for monitoring the evolution of the pandemic over time. This latter type of survey constitutes the proposal of this paper.[9]

c) The third strategy consists of a *specific massive web survey* collected from individuals and households that decide to participate on a voluntary basis. Some 60,000 Israelis completed the online daily survey developed by the Weizmann Institute. The participants disclosed personal details, such as their age, gender, address, general state of health, isolation status and any symptoms they may have been experiencing (Rossman et al., 2020). We observed examples of the same strategy in Iceland, Estonia and other countries. The results allow us to compare contagion and testing experiences for people and households with different socioeconomic characteristics. For strategy a), the self-selection mechanism in the sampling process makes it impossible to extend the results to the whole population.

d) Another possible strategy is to use *pre-existing sample surveys* and partially modify them to collect information about the epidemic. Creating an EU 'Corona Panel', which is a standardized European sample test to uncover the true spread of the coronavirus, is indeed the proposal of the Centre for European Policy Studies, as presented by Daniel Gros (2020). The proposal refers, in particular, to the use of the EU-wide sample of the panel of households that participate in regular surveys on economic and social conditions, called the '*EU statistics on income and living conditions*' (EU-SILC). More specifically, Dewatripont et al. (2020) suggested implementing two tests using the EU-SILC panel: the first aimed at assessing whether the subject is currently infected, and the second aimed at testing whether the person has become immune due to previous exposure.

---

[8] France and Spain are still testing only for case with specific symptoms and for close contacts of infected people.

[9] UK and Italy recently conducted sample surveys at a national level to estimate the real prevalence rate of the infection (ONS, 2020; Istat, 2020). A critical review of the available data on Covid-19 and on the Italian sample survey project is contained in Alleva and Zuliani 2020 and Alleva, 2020.

Timeliness is crucial. In this respect, the latter strategy seems to guarantee good results for the European Statistical System (ESS). A quick reflection could be made on the feasibility of inserting additional modules in the questionnaire of the quarterly Labour Force Survey (LFS), obviously in accordance with the data protection authorities.

The International Labour Organization (ILO) has reached out to the National Statistical Offices (NSOs) to understand the impacts of COVID-19 on their statistical operations, particularly in the domain of labour statistics (ILO, 2020). The ILO recommended that all countries consider what additional information could be useful for capturing the relevant aspects of the epidemic. NSOs should consider whether some existing topics are of low priority; if so, they can thus be temporarily removed from the surveys to create space for new questions.

Many countries are employing combinations of the previously described approaches for collecting data on the epidemic as well integrating them with administrative data or other official statistical sources. While sample surveys represent a bedrock for making inferences about the whole population, planning and building integrated informative systems for the epidemic is certainly the right way to attain a deeper comprehension of the phenomenon. Finally, we observe that in this framework, new data sources (such as mobile phones, web-scraped data, and Internet-of-Things data[10] used to trace the movements of people) should provide useful contributions.

### 3. The basic sampling framework

In what follows, we aim to propose an observational protocol for the estimation of the number of people infected by SARS-CoV-2 (Alleva et al., 2020). Starting from a population where it has been ascertained that individuals are infected (the population contains *verified* cases), the goal is to estimate the portion of the population that is infected but shows no symptoms (the *asymptomatic* cases). For the purpose of the proposed procedure, the individuals are preliminarily classified into two subgroups of interest, which we refer to as *Group A* and *Group B*.

Group A is the subgroup consisting of individuals for which a state of infection has been verified (they could be either hospitalized or in compulsory quarantine) and of all the people who had contact with them in the previous days. Below, we propose to observe the contacts made up to 14 days before the infection has been diagnosed, with this length being the internationally accepted maximum incubation time. However, the unbiasedness of the sampling strategy we propose is still valid (even if less efficient) if the contacts are reconstructed for a shorter time period (e.g., 7 days). Therefore, this group contains all individuals who are foreseen to be infected and not just those for whom their infection status has already been ascertained. Therefore, this group represents both the *apparent* and *latent* dimensions of the epidemic.

---

[10] For example, data collected through images are useful for tracking the movements of people or vehicles or for detecting gatherings in specific places.

Group B contains both healthy people for whom the infection is considered *latent* and those whose infections are still in a phase of incubation, where symptoms can manifest at a future moment in time (up to 14 days later).

The rationale for this breakdown of the population is related to the feasibility of the observational scheme that we propose. Indeed, the proportion of infected people in Group A is much larger than that observed in Group B. Moreover, the number of verified infected people is known through the data collected by health public authorities. Thus, focusing resource investments on observing the contacts of this group maximizes the number of infected people observed in the sample. Nevertheless, it is necessary to observe Group B to produce reliable estimates for the whole population, and this is mandatory for correctly estimating the rate of infected people and the rate of lethality.

Estimates relative to the two subgroups may be obtained on the basis of continuous observations over time and by following two distinct methodologies, both of which based on what is known as *indirect sampling* (Lavalle, 2007; Kiesl, 2016). Indirect sampling is the same technique that is commonly used for the estimation of rare and elusive populations (Sudman, 1988; Thompson and Seber, 1996).

It is important to emphasize that the distinctive element of our proposal lies in the estimate of the infected population obtained by combining the results obtained through two samples drawn from populations A and B. This estimate can establish different roles in relation to the various developmental phases of the epidemic (in terms of the sample size and/or type of diagnostic assessment to be carried out).

At the beginning of the epidemic, the infection has the characteristics of rapidity, unpredictability in terms of the level of spread, and apparent concentration in certain geographical areas and categories of subjects. The response of the health system and the containment measures to be used are not yet codified, nor is the behaviour of the population that should be considered "responsible". In this phase, an investigation strategy based on indirect sampling appears to be coherent, with the strategy starting from the immediate surroundings of subjects who have confirmed infections. This is the sampling strategy proposed for Group A that, in addition to the estimation of a rare phenomenon in the population, also provides an immediate (and continuous over time) response to the epidemic where it explicitly manifests itself.

On the other hand, to measure the intensity and the evolution of the phenomenon for large territorial domains and in general with regard to relevant characteristics of people (gender, age, educational qualification, professional status and more), a traditional population panel survey with sample rotation can be carried out for Group B. The survey is associated with an indirect sampling mechanism so that it can trace and sample the individuals who came into contact with the infected people found in this second sample. This panel survey becomes fundamental during phases that follow the peaks of the epidemic to measure not only the reduction in the number of infections (and therefore to test the positive effects of the containment measures) but also the proportion of the population that had contacts with the virus in the past. During the decline phase of the epidemic (which naturally does not preclude the arrival of new infections in specific territories and environments), the role of the sample from population Group B is fundamental and representative of the entire population followed over time. On the other hand, a diagnostic test must also be identified that takes the relative importance of the infected population and the population susceptible to infection during the various phases of the epidemic into account. From an operational point of view, it seems convenient to rely on nasopharyngeal swabs for sampling the contacts in Group A, regardless of the phase of the epidemic. For the panel survey, a serological examination may be more convenient, particularly during the declining phase in combination with a part of the sample yet to be evaluated (the swab is also administered to this portion[11]).

The combination of the two sampling strategies (with different weights for the ascending and descending phases of the epidemic) represents the competitive advantage of our proposal: it is a dynamic monitoring tool designed to be suitably calibrated both during the growth phase of the infection, providing estimates according to different categories of severity, and during the decline phase, with estimates extended to the parameters of the progressive immunization model for the population.

The advantage of our proposal over a strategy based exclusively on indirect sampling or only on the panel sample can be measured in terms of greater efficiency (and therefore more accurate estimates) and lower investigation costs required to achieve the predetermined levels of precision.

## 4. Specification of the total number of infected people and its breakdown

In what follows, let $U$ be the population of interest of size $N$, and let $k\,(k = 1, \dots, N)$ denote a person belonging to it. Let $v_k$ be a dichotomous variable that assumes a value of 1 if the state of infection is verified and a value of 0 otherwise. Let $U_v = \{k \in U: v_k = 1\}$ be the subpopulation of $U$ for whom the infection is verified and let $U_C = U \setminus U_v$ be the complementary subset.

Let $y_k$ be the value of variable $y$, for person $k$, where $y$ is equal to 1 if the person is infected and 0 otherwise. If $v_k = 1$, then obviously $y_k = 1$; however, if $v_k = 0$, then it is possible that either $y_k = 1$ (an infected person for whom the infection has not yet been verified) or $y_k = 0$ (a healthy person).

The target parameter of our survey, $Y$, is the total number of infected people (verified or not), that is:

$$(1)\ Y = \sum_{k \in U} y_k.$$

Let $l_{k,j}$ be a generic entry of a link matrix ($k=1,2,\dots,N$; $j=1, 2,\dots N$) that is equal to 1 if individual $k$ had contacts with individual $j$ in the past 14 days and 0 otherwise, with $l_{k,k} = 1$ by definition. Starting from $U_v$, it is possible to determine the total number of infections $y$ related to Group A:

---

[11] It is important to emphasize that while the swab allows for an estimation of the infected population at a given moment in time, the serological test allows for the estimation of the portion of the population that had contact with the virus without a time reference. On the other hand, both diagnostic tools provide estimates that are affected by errors, and consequently, the estimates must be considered in probabilistic terms. In particular, to ensure the reliability of the results, while the health protocols for the swab require that the test be repeated over time to ascertain the healing of those who contracted the virus, for the serological examination, diagnostic kits that ensure predetermined levels of specificity and sensitivity can be considered. For a discussion on the impact of these errors in epidemic stages characterized by a different base rates of infection, see Fuggetta (2020).

$$U_A = \left\{ j \in U: \sum_{k \in U_v} l_{k,j} \geq 1 \right\}$$

where $U_A$ includes the subset $U_v$ and all the contacts of the members of that subset. We express this formally as:

$$(2)\ Y_A = \sum_{k \in U_v} \sum_{j \in U} \frac{1}{L_{vj}} l_{k,j} y_j,$$

where

$$(3)\ L_{vj} = \sum_{k \in U_v} l_{k,j}$$

is a quantity introduced to control for the *multiplicity* of the measurement of $y_j$ among the different *k* units of $U_v$ in Equation (2).

On the other hand, starting from $U_C$, it is possible to determine the total number of infections *y* related to *Group B*:

$$U_B = \left\{ j \in U: \sum_{k \in U_C} l_{k,j} \geq 1 \right\}$$

where $U_B$ includes $U_C$ and all the contacts of the infected people in $U_C$:

$$(4)\ Y_B = \sum_{k \in U_C} y_k \sum_{j \in U} \frac{1}{L_{Cj}} l_{k,j} y_j,$$

where, analogously to Equation (3), the quantity

$$(5)\ L_{Cj} = \sum_{k \in U_C} y_k\, l_{k,j}$$

is introduced to control for the *multiplicity* of the measurement of $y_j$ in (4) among the different *k* units in $U_C$.

The sets $U_A$ and $U_B$ can obviously overlap. Let us define their intersection as the set

$$U_{AB} = U_A \cap U_B = \{ j \in U: L_{vj} L_{Cj} \geq 1 \}.$$

The total number of infected people *y* in $U_A \cap U_B$ is given by:

$$(6)\ Y_{AB} = \sum_{j \in U: L_{vj} L_{Cj} \geq 1} y_j.$$

We may obtain alternative expressions of $Y_{AB}$ starting from the sampling frames of $U_v$ and $U_C$:

(7a) $Y_{AB} = \sum_{k \in U_v} \sum_{j \in U} \frac{1}{L_{vj}} l_{k,j} y_j \mathbb{I}(L_{Cj} \geq 1),$

(7b) $Y_{AB} = \sum_{k \in U_C} y_k \sum_{j \in U} \frac{1}{L_{Cj}} l_{k,j} y_j \mathbb{I}(L_{vj} \geq 1)$

where $\mathbb{I}(A)$ equals 1 if $A$ is true and 0 otherwise. The expressions (7a) and (7.b) are useful during the estimation phase, as illustrated in Section 6.3.

Finally, we have

(8) $Y = Y_A + Y_B - Y_{AB}.$

The above setup is illustrated in Figure 1 below.

**Figure 1. Population of interest and its breakdown among the different groups**

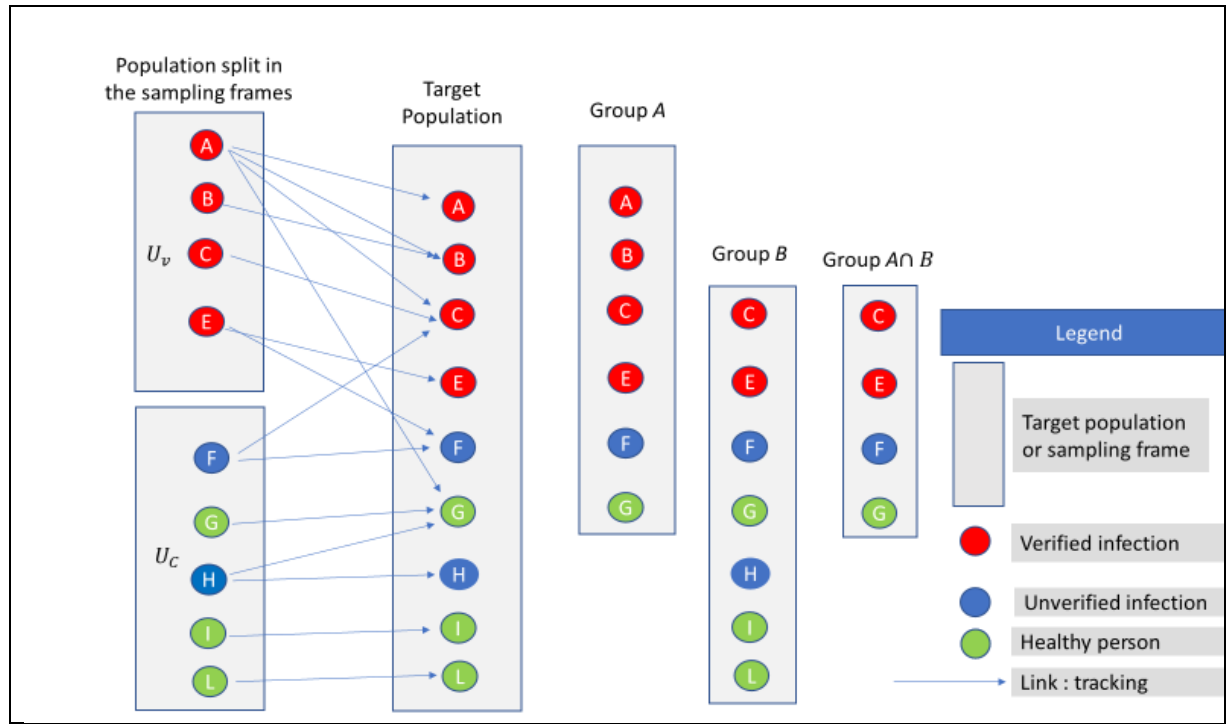

## 5. The sampling design

### 5.1. General description of the sampling schemas

Two independent samples, namely, $S_v$ and $S_C$, are selected from the two population subsets $U_v$ and $U_C$, which represent the sampling frames. The contacts of all infected people in each sample are tracked. The first sample $S_v$ is used to produce an unbiased estimate of $Y_A$, while $S_C$ is used to estimate the total $Y_B$. The total $Y_{AB}$ is estimated from both samples.

### 5.2. Sampling from $U_v$

The subset of the people with verified infections increases over time. It is therefore necessary to set up a sampling mechanism that is realized continuously over time. To simplify the sampling description, let us suppose that $U_v$ represents the set of people with verified infections in a given time period. The sampling of $U_v$ is carried out in the following phases:

    a) A sample $S_v$ is selected without replacement from $U_v$, where the inclusion probabilities are $\pi_{vk}$ $(k = 1, 2 \ldots, \#U_v)$.

    b) All the contacts $U_k = \{j \in U: l_{k,j} = 1\}$ of individual $k$ (selected from $S_v$) are tracked going back 14 days.

    c) A sample $S_{vk}$ is selected from $U_k$ without replacement and with equal probabilities of inclusion $\pi_{2v|k}$. We use the "2" in $\pi_{2v|k}$ to indicate that this is the inclusion probability of the second stage of the sampling process given the selection of person $k$ in the first stage.

At the end of the above process, the sample $S_A = S_v \cup_{k=1}^{\#S_v} S_{vk}$ is formed with an indirect sampling mechanism that includes people from both $S_v$ (verified infected people) and $\cup_{k=1}^{\#S_v} S_{vk}$ (tracked contacts going back 14 days).

The test for verifying an infection is carried out on all the tracked contacts $\cup_{k=1}^{\#S_v} S_{vk}$. Thus, the value of $\psi$ is known for all the people in $S_A$.

**Remark 1**. The process of tracking all the contacts of a person could be complex and cumbersome. Different solutions are possible. One possibility is to leverage digital apps, allowing for epidemic control with digital contact tracing, as suggested by Ferretti et al. (2020). Similarly, Ascani (2020) suggested a method based on personal interviews. In this case, the interviewees must be guided in remembering their contacts by means of a specific structure based on the reconstruction of the "social networks" contacted in the days preceding the infection (Scott, 2000 and Yang et al., 2016).

**Remark 2**. It is clear that for health and wellbeing reasons and to prevent the spread of the infection, it would be best to examine all infected people. However, from a statistical point of view, obtaining high-quality estimates regarding the number of infected persons is not strictly necessary. From this point of view, it is more important to concentrate effort on repeating the examination regularly over time. The effort required to perform a complete study on the whole population would be unsustainable.

#### 5.2.1. Definition of the sampling design

The sampling mechanism for selecting $S_v$ depends on how the data frames for $U_v$ are organized. There are two main possibilities:

**Option 1.**   The data of $U_v$ are available in a centralized data set that can be used for selecting the sample.

**Option 2.**   The data of $U_v$ are available only at a decentralized level so that each healthcare institution has its own list.

The two available options are discussed in turn in the next two subsections.

### 5.2.1.1 Sampling mechanism for Option 1

If the sampling frame of the infected people is centralized in a unified dataset, one could define a *one-stage* design by directly selecting the sample units from the dataset. The selection of the sample can be carried out with the cube algorithm (Deville and Tillé, 2004, 2005), thus ensuring that the Narain Hortvitz-Tompson estimates (Narain, 1951; Horvitz and Thompson, 1952) of the selected sample reproduce the known totals of some auxiliary variables (e.g., distribution by sex and age, employment status, geographical distribution, etc.). This can be expressed as follows:

$$\sum_{k \in S_v} \frac{\mathbf{x}_k}{\pi_{vk}} = \sum_{k \in U_v} \mathbf{x}_k,$$

where $\mathbf{x}_k$ is a vector of $P$ auxiliary variables available for unit $k$.

The definition of the optimal inclusion probabilities $\pi_{vk}$ for indirect sampling that minimize the cost and ensure a predefined level of accuracy for the sampling estimates (or, inversely, minimize the sampling variances for a given budget) can be determined as illustrated by Falorsi and Righi (2019). Tillé and Wilhelm (2017) suggested selecting a sample satisfying Equation (9) through a balanced spatial sampling algorithm that is somehow optimal in maximizing the entropy and minimizing the spatial correlations between neighbouring units (Arbia, 1994; Arbia and Lafratta, 1997, 2002).

Falorsi and Righi (2015) demonstrated that balancing Equation (9) is quite general and allows for the definition of a wide class of sampling designs, including simple random sampling without replacement (SRSWOR), stratified random sampling without replacement (STSRSWOR), stratified random sampling with probability proportional to size (PPS), sampling designs with incomplete stratification (SDIS) and many others.

Assuming that an *SRS* design is used, to obtain the statistical estimates of the number of infected persons in a given *spatial* (the whole national territory or specific geographic area, such as, for example, a region) and *temporal* domain (week/day), it would be sufficient to select approximately 1,000 individuals among the contacts of the infected set of persons for testing. This sample size would ensure a reliable estimate with a sampling error of approximately 5% under the assumption that the proportion of infected people in the target population is approximately 25%.

### 5.2.1.2. Sampling mechanism for Option 2

If the sampling frames for $U_v$ are available only at the healthcare institution level, the selection of units in $S_v$ can be carried out with a two-stage mechanism:

1. **First stage.** A sample $S_{1v}$ of health care institutions is selected from the population of health care institutions (call it $U_{1v}$). The first-stage sample is selected without replacement and with PPS, where healthcare institution $i$ is selected with an inclusion probability given by:

(9) $\pi_{1i} = m \dfrac{M_i}{M},$

in which *m* is the selected number of healthcare institutions to be included in the first-stage sample, $M_i$ is a measure of the size of unit *i* and *M* is the overall measure of size. We may define the measure of size according to different criteria. A good option would be the number of beds available for SARS-CoV-2 patients. The sampling of the health care institutions can be carried out with the already-quoted "cube algorithm", thus ensuring that the Narain Hortvitz-Tompson estimates of the selected first-stage sample reproduce the known characteristics of some auxiliary variables available for the population $U_{1v}$ (e.g., geographical distribution, number of beds available for SARS-CoV-2 patients, etc.). This can be expressed as:

$$(10) \sum_{i \in S_{1v}} \frac{\mathbf{x}_{1v}}{\pi_{1v}} = \sum_{k \in U_{1v}} \mathbf{x}_k,$$

where $\mathbf{x}_{1v}$ is a vector of auxiliary variables for unit *k*. As suggested for Option 1, the sample could be selected (satisfying Equation (10)) with a balanced spatial sampling algorithm that is optimal, maximizes the entropy and minimizes the spatial correlations of the neighbouring units. Even in this case, the balancing of Equation (10) allows us to define the general class of sampling designs described by Falorsi and Righi (2015).

2. **Second stage.** A fixed number, e.g., $\bar{n}$, of infected people is selected from the sampled institution by *drawing the units* without replacement via a simple random sampling procedure.

In such a way, the sampling process is *self-weighting* (Murthy and Sethy, 1965) in the sense that all the units in $U_v$ have an equal probability of being selected. Indeed, the final inclusion probability of person *k* being selected from healthcare institution *i* is given by the following expression:

$$(11) \pi_{vk} = m \frac{M_i}{M} \frac{\bar{n}}{M_i} = m \frac{\bar{n}}{M}.$$

The *self-weighting* property defines a sampling design that is somehow optimal (Kish, 1966) in the sense that it avoids the negative impact of the variability of the sampling weights on the sampling variances.

The sampling selection criterion could be based on a time mechanism, as this is feasible and easily implementable at a decentralized level. For instance, a sample of infected people could be selected by considering those who had access to the healthcare institution within a two-hour time period.

### 5.3. Sampling from $U_C$

In this section, we illustrate the sampling design for the first selection process, where we sample a panel of individuals independently from $S_v$ for estimating the total $Y_B$. Afterwards, we monitor these people repeatedly over time.

The operational aspects to be carried out in this first sampling process are as follows:

a) First, a sample $S_C$ is selected without replacement from $U_C$, where the inclusion probabilities are $\pi_{Ck}$ ($k = 1, 2 \ldots, \#U_C$).

b) The people in the panel take a diagnostic test on a regular basis (for example, once a month). If member $k$ of the panel receives a positive test result (i.e., $y_k = 1$), all their contacts $U_k$ are tracked up to 14 days back in time.

c) If $y_k = 1$, a sample $S_{Ck}$ is selected from $U_k$ without replacement and with equal inclusion probability $\pi_{2C|k}$. We adopt $\pi_{2C|k}$ for the second stage inclusion probability, where the same notation as that of $\pi_{2v|k}$ is used. At the end of the whole process, the sample $S_B = S_C \cup_{k=1: y_k=1}^{\#S_C} S_{Ck}$ is formed with an indirect sampling mechanism, including people from both $S_C$ (people for whom their infection statuses are not known) and $\cup_{k=1: y_k=1}^{\#S_C} S_{Ck}$ (tracked contacts of the infected people in $S_C$, going back 14 days).

**Remark 3**. The populations $U_v$ and $U_C$ change as a function of time. The panel can be representative of the shifting population. We discuss this topic later on in Section 7. Here, we note that in the subsequent surveys, the verified infected people in the panel are automatically captured by the sampling mechanism defined for the population $U_v$. However, sample $S_C$ is smaller in size than the total population, observing only the non-verified infected people. This reduction in the sampling size makes it necessary to regularly refresh the panel over time.

### 5.3.1. A note on some practicalities of the sampling design

The sampling design of the panel can be carried out according to different schemas, depending on the availability of the frame and on other organizational aspects. One possibility is to form a subsample from a regular survey of households carried out by official statistics. Here, we assume that the frame of $U$ is represented by a register that is available at a central level and that for each sample unit, we form a set of auxiliary variables. Furthermore, we assume that in this register, the subset $U_C$ can also be identified.

In this informative context, a one-stage sampling design can be carried out with optimal inclusion probabilities $\pi_{Ck}$, as determined following the steps of Falorsi and Righi (2015, 2019). The sampling can then be carried out with a balanced spatial sampling algorithm (Tillé and Wilhelm, 2017), thereby ensuring that the following balancing equations are satisfied:

$$(12) \quad \sum_{k \in S_C} \frac{\mathbf{x}_k}{\pi_{Ck}} = \sum_{k \in U_C} \mathbf{x}_k.$$

**Remark 4.** The *panel* can be constructed using a two-phase design so that the selection process can be executed by *pre-screening* two sub-groups, namely:

- A number of individuals who continue to travel (and are therefore more subject to being infected than individuals who are not traveling).
- A number of individuals with few contacts who fully observe the prescribed quarantine recommendations.

**Remark 5.** The two-phase mechanism could be useful if the identification of $U_C$ cannot not be carried out. This can be realized in the two-step pre-screening phase.

The number of persons involved in the *panel* may be approximately 1,000 (to obtain approximately 1,200 tested individuals) for a given *territorial* and *temporal* sampling domain, thus guaranteeing a reliable estimation with a sampling error of approximately 10% (assuming that the proportion of infected people in this target population is approximately 10%).

### 5.4. Final comments on the sampling design

We first note that in our proposal, we subsample from the determined list of contacts. We adopt this choice for controlling the costs associated with the survey. However, we could extend the sample to all sets of contacts. Furthermore, if we continue tracking the contacts until all the people being tracked are not infected, the adopted sampling design becomes a classic adaptive schema (Thompson and Seeber, 1996), which can thus be seen as a particular application of our proposal.

Given the complexity of the epidemiology of COVID-19, it may be useful to consider subgroups in Group B. This may become useful based on the need to consider heterogeneous models (i.e., considering heterogeneous populations), as this type of model seems to be required for the infectious agent. In particular, it may be important to consider breaking down certain epidemiological parameters into different subgroups (e.g., transmission coefficient, time to become infectious, proportion of detected cases, time from infection to detection, time to recover). Therefore, we suggest defining 4 subgroups considering two binary factors: low-risk/high-risk and low-mobility/high-mobility. These subgroups are described as follows:

- A number of individuals not belonging to high-risk groups who continue to travel/work (and are therefore more subject to being infected and infectious than non-travellers).
- A number of people not belonging to high-risk groups with few contacts who fully observe the prescribed quarantine recommendations.
- A number of individuals belonging to high-risk groups who continue to travel/work (and are therefore more subject to being infected and infectious than non-travellers), such as health-care workers.
- A number of people belonging to high-risk groups with few contacts who fully observe the prescribed quarantine recommendations.

For Group A, there might be some advantage in considering the same 4 subgroups, since the transmission coefficient of each of these subgroups can be significantly different from the others.

Considering 4 subgroups in both Groups A and B may impact the sample size required to obtain a given sampling error at the subgroup level. Group B has the potential for enabling the study of some crucial "invisible" parameters of the epidemiology of COVID-19 (e.g., the proportion of asymptomatic cases, the time for symptomatic and asymptomatic people to become infectious, and even the proportion of undetected symptomatic cases) in detail. This is also true for each of the 4 subgroups independently. The sample size for Group B should be defined with this in mind. Population density is also an important factor to control when designing the sampling process.

## 6. Sample estimation of the total number of infected people

We can compute a direct estimation of the total number of infected people $Y$ for each time and each territorial unit as:

$$(13) \quad \hat{Y} = \hat{Y}_A + \hat{Y}_B - \hat{Y}_{AB},$$

with

$$(14) \quad \hat{Y}_{AB} = \alpha \hat{Y}_{AB}^A + (1 - \alpha)\hat{Y}_{AB}^B,$$

where $\hat{Y}_A$ and $\hat{Y}_{AB}^A$ are the generalized weight share method (GWSM, Lavallé, 2007) estimates of the totals $Y_A$ and $Y_{AB}$ derived from the sample $S_A$; $\hat{Y}_B$ and $\hat{Y}_{AB}^B$ are the GWSM estimates of the totals $Y_B$ and $Y_{AB}$ calculated from the sample $S_B$; and $\hat{Y}_{AB}$ is a convex combination of the GWSM estimates $\hat{Y}_{AB}^A$ and $\hat{Y}_{AB}^B$, with $0 \leq \alpha \leq 1$. The parameter $\alpha$ can either be fixed in advance or calculated from the survey data. Further discussion on the choice of $\alpha$ is provided in Section 6.3.

### 6.1. Estimation of the component $\hat{Y}_A$

The GWSM estimator of the total number of infected people in group $A$, as expressed in Equation (2), is given by:

$$(15) \quad \hat{Y}_A = \sum_{k \in S_v} \frac{1}{\pi_{vk}} \sum_{j \in S_{vk}} \frac{1}{\pi_{2v|k}} \frac{1}{L_{vj}} l_{k,j} y_j$$

$$= \sum_{k \in S_v} \frac{1}{\pi_{vk}} \hat{Z}_{vk},$$

where

$$(16) \quad \hat{Z}_{vk} = \sum_{j \in S_{vk}} \frac{1}{\pi_{2v|k}} \frac{1}{L_{vj}} l_{k,j} y_j$$

represents the second-stage estimate of

$$(17) \quad Z_{vk} = \sum_{j \in U_k} \frac{1}{L_{vj}} l_{k,j} y_j.$$

**Remark 6.** The term $L_{vj}$ in the previous equation corresponds to the total number of contacts of unit $j$ with people who have verified infections. It can be collected either with digital contact tracing (Ferretti, 2020) or by interviews.

*Proof of the unbiasedness of $\hat{Y}_A$*

This proof can be found in Section 5.1 of Lavallée (2007). Denoting the sampling expectation operator as $E(\cdot)$, we have

$$(18)\ E(\hat{Y}_A) = E\left[\sum_{k \in U_v} \sum_{j \in U} \frac{\delta_{vk}}{\pi_{vk}\pi_{2v|k}} \frac{\delta_{2vj|k}}{L_{vj}} l_{k,j} y_j\right],$$

where $\delta_{vk}$ is a dichotomous variable with $\delta_{vk} = 1$ if $k \in S_v$ and $\delta_{vk} =$ otherwise. $\delta_{2vj|k}$ is a second dichotomous variable with $\delta_{2vj|k} = 1$ if $j \in S_{vk}$ and 0 otherwise.

From Equation (18), we obtain:

$$(19)\ E(\hat{Y}_A) = \sum_{k \in U_v} \sum_{j \in U} \frac{E(\delta_{vk}\delta_{2vj|k})}{\pi_{vk}\pi_{2v|k}} \frac{1}{L_{vj}} l_{k,j} y_j.$$

However, since:

$$(20)\ E(\delta_{vk}\delta_{2vj|k}) = E[\delta_{vk} E(\delta_{2vj|k}|\delta_{vk} = 1)] = E[\delta_{vk}\pi_{2v|k}] = \pi_{vk}\pi_{2v|k},$$

plugging Equation (20) into Equation (19), we finally obtain:

$$E(\hat{Y}_A) = \sum_{k \in U_v} \sum_{j \in U} \frac{1}{L_{vj}} l_{k,j} y_j = Y_A. \hspace{2cm} \text{Q. E. D.}$$

### Variance of $\hat{Y}_A$

The main results on this topic can also be found in Section 5.1 of Lavallée (2007). On the basis of the theorem on two-stage sampling (Cochran, 1977), the variance of $\hat{Y}_A$ can be expressed as follows:

$$(21)\ V(\hat{Y}_A) = V_1\left(\sum_{k \in S_v} \frac{1}{\pi_{vk}} Z_{vk}\right) + \sum_{k \in U_v} \frac{1}{\pi_{vk}} V_2\left(\sum_{j \in S_{vk}} \frac{1}{\pi_{2v|k}} \frac{1}{L_{vj}} l_{k,j} y_j,\right).$$

In the previous expression, the variance is decomposed into the sum of the first-stage variance ($V_1$) and the first-stage expectation of the second-stage variance ($V_2$). All the elements of the previous expression can be estimated with standard statistical inferential techniques (see Horvitz and Thompson, 1952 and Kish, 1965).

### 6.2. Estimation of the component $\hat{Y}_B$

The GWSM estimator of the component $\hat{Y}_B$ is given by:

$$(22)\ \hat{Y}_B = \sum_{k \in S_C} \frac{1}{\pi_{Ck}} y_k \sum_{j \in S_{Ck}} \frac{1}{\pi_{2C|k}} \frac{1}{L_{Cj}} l_{k,j} y_j$$

$$= \sum_{k \in S_C} \frac{1}{\pi_{Ck}} \hat{Z}_{Ck}$$

where the term

$$(23)\ \hat{Z}_{Ck} = y_k \sum_{j \in S_{Ck}} \frac{1}{\pi_{2C|k}} \frac{1}{L_{Cj}} l_{k,j} y_j$$

represents the estimate of

$$(24)\ Z_{Ck} = y_k \sum_{j \in U_k} \frac{1}{L_{Cj}} l_{k,j} y_j.$$

### *Proof of the unbiasedness of $\hat{Y}_B$*

To prove the unbiasedness of $\hat{Y}_B$, we start with:

$$(25)\ E(\hat{Y}_B) = \sum_{k \in U_C} y_k \sum_{j \in U_k} \frac{E(\delta_{Ck}\, \delta_{2Cj|k})}{\pi_{Ck}\, \pi_{2C|k}} \frac{1}{L_{Cj}} l_{k,j} y_j,$$

where $\delta_{Ck}$ is a dichotomous variable with $\delta_{Ck} = 1$ if $k \in S_C$ and $\delta_{Ck} = 0$ otherwise. $\delta_{2Cj|k}$ is a dichotomous variable with $\delta_{2Cj|k} = 1$ if $y_k = 1 \cap j \in S_{Ck}$ and 0 otherwise.

However, we have:

$$(26)\ E(\delta_{Ck}\, \delta_{2Cj|k}) = E[\delta_{Ck} E(\delta_{2Cj|k}|\delta_{Ck} = 1)] = E[\delta_{Ck} \pi_{2C|k}] = \pi_{Ck}\, \pi_{2C|k}.$$

From Equations (25) and (26) it follows that:

$$E(\hat{Y}_B) = \sum_{k \in U_C} y_k \sum_{j \in U} \frac{1}{L_{Cj}} l_{k,j} y_j. \qquad \text{Q. E. D.}$$

The term $L_{Cj}$ corresponds to the total number of contacts of unit *j* with people who have unverified infections. Similar to the estimation process of $\hat{Y}_B$, this information can be collected either with digital contact tracing or by interviews. Alternatively, we can determine $L_{Cj}$ by following a *back-tracing process*: if unit *j* is infected, we should test the all their contacts for COVID-19.

## Variance of $\hat{Y}_B$

The variance may be obtained by simply adapting expression (21):

$$(27)\ V(\hat{Y}_B) = V_1\left(\sum_{k\in S_C}\frac{1}{\pi_{Ck}}Z_{Ck}\right) + \sum_{k\in U_C}\frac{1}{\pi_{Ck}}V_2\left(y_k\sum_{j\in S_{Ck}}\frac{1}{\pi_{2C|k}}\frac{1}{L_{Cj}}l_{k,j}y_j\right).$$

### 6.3. Estimation of the component $\hat{Y}_{AB}$

Starting from expression (7a), by using the data from sample $S_A$, we obtain the GWSM unbiased estimator of $Y_{AB}$ as:

$$(28)\ \hat{Y}_{AB}^A = \sum_{k\in S_v}\frac{1}{\pi_{vk}}\sum_{j\in S_{vk}}\frac{1}{\pi_{2v|k}}\frac{1}{L_{vj}}l_{k,j}\,y_j\,\mathbb{I}(L_{Cj}\geq 1).$$

Starting from expression (7b), by using the data from sample $S_B$, we derive the GWSM unbiased estimator of $Y_{AB}$ as:

$$(29)\ \hat{Y}_{AB}^B = \sum_{k\in S_C}\frac{1}{\pi_{Ck}}\sum_{j\in S_C}\frac{1}{\pi_{2C|k}}\frac{1}{L_{Cj}}l_{k,j}y_j\,\mathbb{I}(L_{vj}\geq 1).$$

The information about the intersection of the samples with the subpopulation $U_{AB}$ may be collected either during the interviews or with digital contact tracing.

Singh and Mecatti (2011) provided an in-depth illustration of the different approaches in the literature that are used find the optimal value of $\alpha$ in the context of multiple frame surveys. Hartley (1962, 1974) proposed choosing $\alpha$ in (14) to minimize the variance of $\hat{Y}$. Because the frames are sampled independently, the variance of $\hat{Y}$ is:

$$(31)\ V(\hat{Y}) = V(\hat{Y}_A) + V(\hat{Y}_B) + \alpha^2\,V(\hat{Y}_{AB}^A) + (1-\alpha)^2 V(\hat{Y}_{AB}^B) +$$
$$- 2\alpha Cov\,(\hat{Y}_{AB}^A, \hat{Y}_A) - 2(1-\alpha)\,Cov\,(\hat{Y}_{AB}^B, \hat{Y}_B).$$

Thus, for general survey designs, the variance-minimizing value of $\alpha$ is:

$$(32)\ \alpha^{opt} = \frac{V(\hat{Y}_B) + Cov\,(\hat{Y}_{AB}^B, \hat{Y}_B) - Cov\,(\hat{Y}_{AB}^A, \hat{Y}_A)}{V(\hat{Y}_A) + V(\hat{Y}_B)}.$$

Unfortunately, the above quantity depends on the variable *y*.

Note that if one of the covariances in (32) is large, it is possible for $\alpha^{opt}$ to be smaller than 0 or greater than 1. Hartley (1974) suggested opting for this alternative expression:

$$(33)\ \alpha^* = \frac{V(\hat{Y}_B)}{V(\hat{Y}_A) + V(\hat{Y}_B)}.$$

*Unbiasedness and variance.* The proof of unbiasedness and the calculation of the variance of the estimator $\hat{Y}_{AB}$ are straightforward extensions of what has been illustrated in Sections 6.1 and 6.2.

**Remark 7.** Lavallé and Rivest (2012) proposed estimating the total *Y* with the *generalized capture-recapture estimator* (GCRE), which makes joint use of the capture-recapture *Petersen* estimator and GWSM estimators. In our context, the GCRE estimator may be expressed as:

$$(34)\ \hat{Y}_{GCRE} = \frac{\hat{Y}_A \hat{Y}_B}{\hat{Y}_{AB(S_A \cap S_B)}},$$

where $\hat{Y}_{AB(S_A \cap S_B)}$ is the estimate of $Y_{AB}$ that is computed on the basis of the units observed in the intersection $S_A \cap S_B$. The sampling weights for producing the estimates from $S_A \cap S_B$ are given in formula (11) in the abovementioned paper. With respect to expression (28), the GCRE estimator allows for estimating the hidden population that would not be visible with either the public health structure or with the panel survey (e.g., the people who died at home), and this group is very difficult to capture with the usual survey techniques. The main problem for adopting the GCRE estimator is that it would require an overlap of the samples of Groups A and B.

**Remark 8.** In Section 8 and in the Appendix, we see that the maximum efficiency is achieved by sampling from $U_v$. At the same time, collecting the value of the variable $L_{Cj}$ could be complex due to the need to set up a *back-tracing process*. Thus, a feasible alternative strategy for the estimation of *Y* could be represented by:

$$\hat{Y}_{alt} = \hat{Y}_A + \hat{Y}_C - \hat{Y}_{AC}^A,$$

where $\hat{Y}_C$ is the standard Narain-HT estimate of the total $\psi$ in $U_C$ and $\hat{Y}_{AC}^A$ is the GWSM estimate of the total $\psi$ in the intersection of $U_A$ with $U_C$ obtained by the sample $S_A$. These terms are as follows:

$$\hat{Y}_C = \sum_{k \in S_C} \frac{1}{\pi_{Ck}} y_k,$$

$$\hat{Y}_{AC}^A = \sum_{k \in S_v} \frac{1}{\pi_{vk}} \sum_{j \in S_{vk}} \frac{1}{\pi_{2v|k}} \frac{1}{L_{vj}} l_{k,j}\, y_j \mathbb{I}\big[(L_j - L_{Cj}) \geq 1\big],$$

where $L_j$ is the total number of contacts of unit *j*.

## 7. Sampling design for follow-ups of the survey in subsequent waves

The observational scheme proposed in the above sections is set up as a cross-sectional survey. However, it can be adapted for monitoring the evolution of the number of infected people over time; this is done according to a mechanism that is updated like a chain mechanism time after time. While an in-depth study of this aspect deserves a separate study, we limit ourselves here to introducing the topic and to providing some initial indications.

Let us consider two consecutive points in time, e.g., $t = 0$ and $t = 1$.

Assume person $k$ is verified as infected at time 0 and is hence denoted as $v_{0,k} = 1$. This person may still be infected ($v_{1,k} = y_{1,k} = 1$), or she/he may no longer be infected ($y_{1,k} = 0$) because of *death* (denoted by the dichotomous variable $d_{1,k} = 1$) or *healing* (denoted by the dichotomous variable $h_{1,k} = 1$).

The total of the $y$ variable at time 1 may then be defined as:

(35) $Y_1 = Y_0 + \Delta D_{0\to 1} + \Delta H_{0\to 1} + \Delta Y_1$,

where $Y_0$ is the total number of infections at time 0 and:

(36) $\Delta D_{0\to 1} = \sum_{k\in U} y_{0,k}\, d_{1,k}, \quad \Delta H_{0\to 1} = \sum_{k\in U} y_{0,k}\, h_{1,k}, \quad \Delta Y_1 = \sum_{k\in U} (1 - y_{0,k})\, y_{1,k}.$

In Equation (36), the quantity $(Y_0 + \Delta D_{0\to 1} + \Delta H_{0\to 1})$ indicates the total number of verified infected people at time 0 who are still infected at time 1, while the quantity $\Delta Y_1$ denotes the total number of *new* infections.

The updating of the sampling structures illustrated in the previous sections allows us to obtain a direct estimate of each of the components of (35), as illustrated in Figure 2.

The total $\Delta Y_1$ can be estimated, as described in Section 5, using two sources of data, namely:

- The sample $S_{1,v}$, which automatically captures the new entrants into the verified infected population at time 1. These new entrants are denoted by $\Delta U_{1,v}$, since the sampling selection is carried out on this population continuously over time. Then, a sample of their contacts can obtained as described in Section 4.2, resulting in the sample $S_{1,A}$.

- The panel $S_{0,C}$, which is selected at time $t = 0$ and is updated over time, since the tests carried out at time $t = 1$ on the individuals of $S_{0,C}$ distinguish the *newly infected* people of the panel. Then, tracking the contacts of the infected people allows us to obtain the sample $S_{1,B}$.

The estimation of the totals $(Y_0 + \Delta D_{0\to 1} + \Delta H_{0\to 1})$ can be obtained by following up on the health statuses of the infected people captured by the samples $S_{0,A}$ and $S_{0,B}$ at time 0. The estimates are then obtained with the sampling weights computed at time 0.

Therefore, we have:

(37) $\hat{Y}_1 = \hat{Y}_0 + \widehat{\Delta D}_{0\to 1} + \widehat{\Delta H}_{0\to 1} + \widehat{\Delta Y}_1$,

where $\hat{Y}_0, \widehat{\Delta D}_{0\to 1}, \widehat{\Delta H}_{0\to 1}, \widehat{\Delta Y}_1$ are the direct estimates of the quantities $Y_0, \Delta D_{0\to 1}, \Delta H_{0\to 1}, \Delta Y_1$, respectively. The above mechanism can be updated in a chain mechanism, and thus the estimate for any time $t > 1$ is obtained as:

(38) $\hat{Y}_t = \hat{Y}_{t-1} + \widehat{\Delta D}_{t-1\to t} + \widehat{\Delta H}_{t-1\to t} + \widehat{\Delta Y}_t.$

**Figure. 2.** Follow-up of samples over time

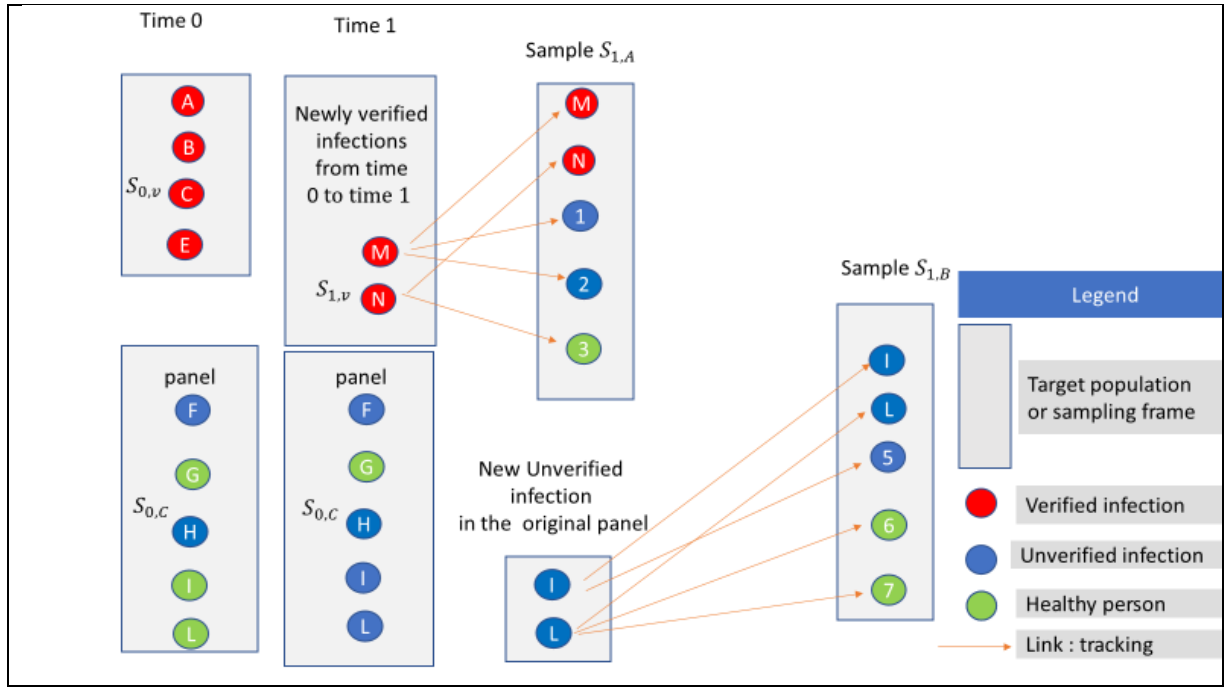

## 8. A note on the efficiency of the proposed strategy

To derive the efficiency of the estimators, we need to specify different cases that may occur, where these cases related to the intersection between the samples from population groups A and B. Here, we consider only two rather realistic cases, which are illustrated in Figure 3.

- **Case 1.** The samples from Groups A and B have the same size, and there is a strong intersection between the two groups. This case could characterize the situation in which there is no control of the infection. In this case, the proportions $\gamma_A = Y_{AB}/Y_A$ and $\gamma_B = Y_{AB}/Y_B$ are slightly smaller than 1. Thus, it is possible to consider $\gamma_A \cong 1$ and $\gamma_B \cong 1$.

- **Case 2.** The sample from Group A is much smaller than the sample from Group B, and there is a strong intersection between the sample from Group A and the intersection between the two samples. In this case, we can consider $\gamma_A \cong 1$ and $\gamma_B \ll 1$. This case could be that in which the infection is controlled by locking down the people.

Figure. 3. Two realistic cases of intersections between the samples from Groups A and B

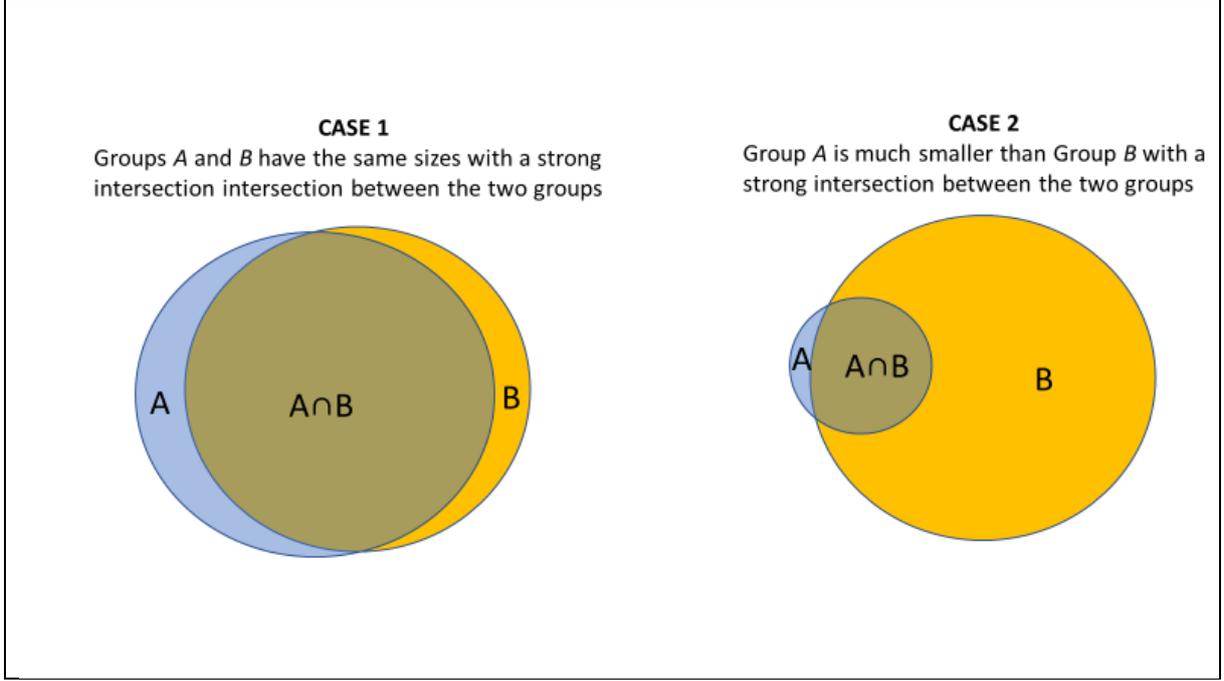

Below, we summarize a result from simple random sampling that may be useful to better understand the efficiency of the strategy proposed here. We give all the details in the Appendix.

Let us suppose that a sample $S$ of size $n$ is selected from $U$ with a SRSWOR design. Let

$$(39)\ \hat{Y}_{HT,SRS} = \sum_{k=1}^{n} \frac{N}{n} y_k = \sum_{k=1}^{n} \frac{1}{f} y_k$$

be the $HT$ estimate of $Y$, where $f = n/N$. Let $\mu = Y/N$ be the proportion of infected people in the overall population.

For a large $N$ and small $f$, the anticipated variance (AV) of $\hat{Y}_{HT,SRS}$ can be approximated by (Falorsi and Righi, 1915, Appendix 4):

$$(40)\ AV(\hat{Y}_{HT,SRS}) = \frac{N}{f} \mu(1-\mu).$$

Let $U_{yl} = \{k,j: y_k = 1, y_k l_{k,j} = 1; k, j = 1, ..., N\}$ denote the sub-population of infected people and those who have had contact with them. Let $\vartheta = Y/\#U_{yl}$ be the proportion of infected people in $U_{yl}$, with $\mu \ll \vartheta$. Let us suppose that the sample $S$ is allocated proportionally between the two frames $U_v$ and $U_C$ and that a SRSWOR is performed in each frame. Thus, the sample sizes of $U_v$ and $U_C$ are $P_v n$ and $(1-P_v)n$, respectively.

The GWSM estimates of the totals $Y_A$, $Y_B$, $Y_{AB}$ and $Y$ are:

$$(41)\ \hat{Y}_{A,SRS} = \sum_{k=1}^{P_v n} \frac{1}{f} \sum_{j=1}^{N} \frac{1}{L_{vj}} l_{k,j}\, y_j,\ \hat{Y}_{B,SRS} = \sum_{k=1}^{(1-P_v)n} \frac{1}{f} y_k \sum_{j=1}^{N} \frac{1}{L_{Cj}} l_{k,j}\, y_j,$$

$$\hat{Y}^A_{AB,SRS} = \sum_{k=1}^{P_v n} \frac{1}{f} \sum_{j=1}^{N} \frac{1}{L_{vj}} l_{k,j}\, y_j\, (L_{Cj} \geq 1),$$

$$\hat{Y}_{AB,SRS}^{B} = \sum_{k=1}^{(1-P_v)n} \frac{1}{f} y_k \sum_{j=1}^{N} \frac{1}{L_{Cj}} l_{k,j} \, y_j \, (L_{vj} \geq 1).$$

$$\hat{Y}_{SRS} = \hat{Y}_{A,SRS} - \alpha \hat{Y}_{AB,SRS}^{A} + \hat{Y}_{B,SRS} - (1-\alpha)\hat{Y}_{AB,SRS}^{B}.$$

Assuming that the number of contacts $L$ is roughly constant in $U$, the AV of $\hat{Y}_{SRS}$ is

(42) $AV(\hat{Y}_{SRS}) = AV(\hat{Y}_{A,SRS} - \alpha \hat{Y}_{AB,SRS}^{A}) + AV[\hat{Y}_{B,SRS} - (1-\alpha)\hat{Y}_{AB,SRS}^{B}]$

where

(43) $AV(\hat{Y}_{A,SRS} - \alpha \hat{Y}_{AB,SRS}^{A}) \cong \frac{P_v N}{f} \frac{1}{L} \vartheta[(1-\vartheta)(1-2\alpha\gamma_A) + \alpha^2 \gamma_A(1-\gamma_A \vartheta)]$

and

(44) $AV(\hat{Y}_{B,SRS} - (1-\alpha)\hat{Y}_{AB,SRS}^{A})$
$\cong (1-P_v)N \frac{1}{fL\mu} \vartheta[(1-\mu\vartheta) + (1-\alpha)^2 \gamma_B(1-\gamma_B \mu\vartheta) - 2(1-\alpha)\gamma_B(1-\mu\vartheta)].$

Comparing expressions (42) and (40), we can see that the efficiency of the proposed strategy can be defined as the ratio of the two AVs:

(45) $Eff(\hat{Y}_{SRS}) = \frac{AV(\hat{Y}_{SRS})}{AV(\hat{Y}_{HT,SRS})}$

$= \frac{\frac{1}{L}\vartheta[(1-\vartheta)(1-2\alpha\gamma_A) + \alpha^2 \gamma_A(1-\gamma_A \vartheta)]}{\mu(1-\mu)} P_v +$

(46) $+ \frac{\frac{1}{L\mu}\vartheta[(1-\mu\vartheta) + (1-\alpha)^2 \gamma_B(1-\gamma_B \mu\vartheta) - 2(1-\alpha)\gamma_B(1-\mu\vartheta)]}{\mu(1-\mu)} (1-P_v).$

Looking at expression (46), we can highlight the following results:

- The effectiveness of the strategy is at its maximum in case 1.
- The efficacy is at its maximum when sampling from $U_v$, for which it is realistic to have:
$$\frac{1}{L}\vartheta[(1-\vartheta)(1-2\alpha\gamma_A) + \alpha^2 \gamma_A(1-\gamma_A \vartheta)] < \mu(1-\mu).$$
- The efficacy could be lower than above or null when sampling from $U_C$ for the case in which the condition
$$\frac{1}{L\mu}\vartheta[(1-\mu\vartheta) + (1-\alpha)^2 \gamma_B(1-\gamma_B \mu\vartheta) - 2(1-\alpha)\gamma_B(1-\mu\vartheta)] < \mu(1-\mu)$$
  is not always given.

Thus, a good strategy could be to oversample from $U_v$ and obtain a small sample from $U_C$.

## 9. Empirical evaluations of the proposed method: a Monte Carlo study
### 9.1 Artificial data generation
Since it is not possible at this stage to include a numerical illustration using real-life sample data, in this section, we report the results of a series of Monte Carlo experiments that numerically justify our proposed ideas and show their statistical performances in an artificial, although as realistic as possible, context.

Before showing our simulation results, we need to clarify the criteria we used in the data generation process and those employed in the generation of the geographical map on which the data are observed. This second element is essential given the peculiar nature of the transmission mechanism, which requires physical proximity between infected people. First, to simulate an artificial population describing the time evolution of an epidemic, we considered a popular model constituted by a system of six differential equations that, at each moment in time, describe six categories of individuals, namely, susceptible people (S), those exposed to the virus (E), those infected with symptoms (I), those without symptoms (A) and those that are removed from the population either because they healed (R) or are dead (D). This modelling framework is a result of the seminal contribution of Hamer (1906), Kermack and McKendrick (1927) and Soper (1929), and it is often referred to as the "SIR model" due to the initials of the categories considered in the first simplified formulation: Susceptibles, Infected and Removed. A comprehensive overview of this model is contained in Cliff et al. (1981). See also Vynnycky and White (2010). Figure 4 diagrammatically describes the transitions between the 6 categories. For the random data generation process, we assumed that if infected, a susceptible person in the population (S) would remain in the exposed state (E) for 5 days. After that period, the subject could either become infected with symptoms (I) with probability 0.25 or without symptoms (asymptomatic; symbol A) with probability 0.75. An asymptomatic person remains infected (and so is still able to transmit the virus) for 14 days. After this period, all asymptomatic patients are considered healed and pass to the "removed" category (R). In contrast, the infected people showing symptoms heal with a probability of 0.85 or die (D) with a probability of 0.15 (death rate case).

Figure 4. The six basic categories of our simulation model and their transition patterns.

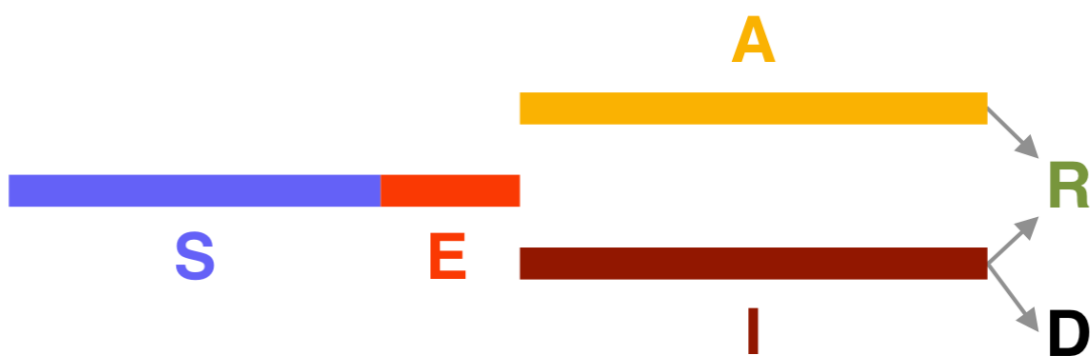

For the map generation process, we considered a population distributed across 25 spatial units laid on a regular 5-by-5 square lattice grid. Each square of the grid contains a number of individuals randomly drawn from a uniform distribution ranging between 800 and 1,000. After performing a simulation exercise with these parameter values, we obtained an artificial population with a total of 22,217 individuals.

This geographical representation is very general in that the map generated in this way can represent, e.g., a city divided into blocks, a region divided into smaller spatial unions, or any other meaningful geographical partition.

The contagion mechanism is favoured for studying human mobility. In our exercise, we assumed that at any moment in time, a certain percentage *m* of the population could move between the squares. We distinguished two epidemic phases. In Phase 1, people are free to move, and this percentage is *m=0.03,* while Phase 2 describes a period of lockdown when mobility is discouraged and *m=0.01.* In particular, we considered Phase 1 as a period of 4 weeks and Phase 2 as the period containing the 8 subsequent weeks. Communication during the lockdown period is limited not only by the number of people who move but also by the extent of their movements. This is a further simulation parameter that is generated by a uniform distribution ranging from -4 to 4 during Phase 1 (thus allowing movements in and out of the cells) and between -1 and 1 during Phase 2. Given the mobility pattern described above, contagion is determined by social interactions and contact opportunities. The number of contacts in each square of the grid is assumed to be determined by a random number drawn from a Poisson distribution with a parameter, i.e., $c_n$, while the number of people involved in the movements is also a Poisson number characterised by a different parameter $c_p$. Given these assumptions, contagion occurs in the following way. If in a meeting at least one asymptomatic or exposed person is present, $i_m$ susceptible people are infected and are moved into the "exposed" category. In our runs of the simulation, we considered Phase 1 to be characterised by the following parameters: *$c_n = 20$; $c_p = 5$; $i_m = 3$.* In contrast, during Phase 2, the three parameters became *$c_n = 3$, $c_p = 3$, and $i_m = 2$,* reflecting the decreased chances of contact between people. Figure 2 describes the time evolutions of the six categories of people in our simulated epidemics. As already stated, we considered Phase 1 to include 4 weeks (day 1 to day 28) and Phase 2 to include 8 weeks (day 29 to day 84). Figure 5 shows that despite the many assumptions that we were forced to include in the simulation, the contagion curves are very similar to those observed worldwide in the recent 2020 SARS-CoV-2 pandemic.

**Figure 5. Time evolutions of the six categories of people in the simulated epidemics. Phase 1 refers to days 1- 28. Phase 2 refers to days 29-84.**

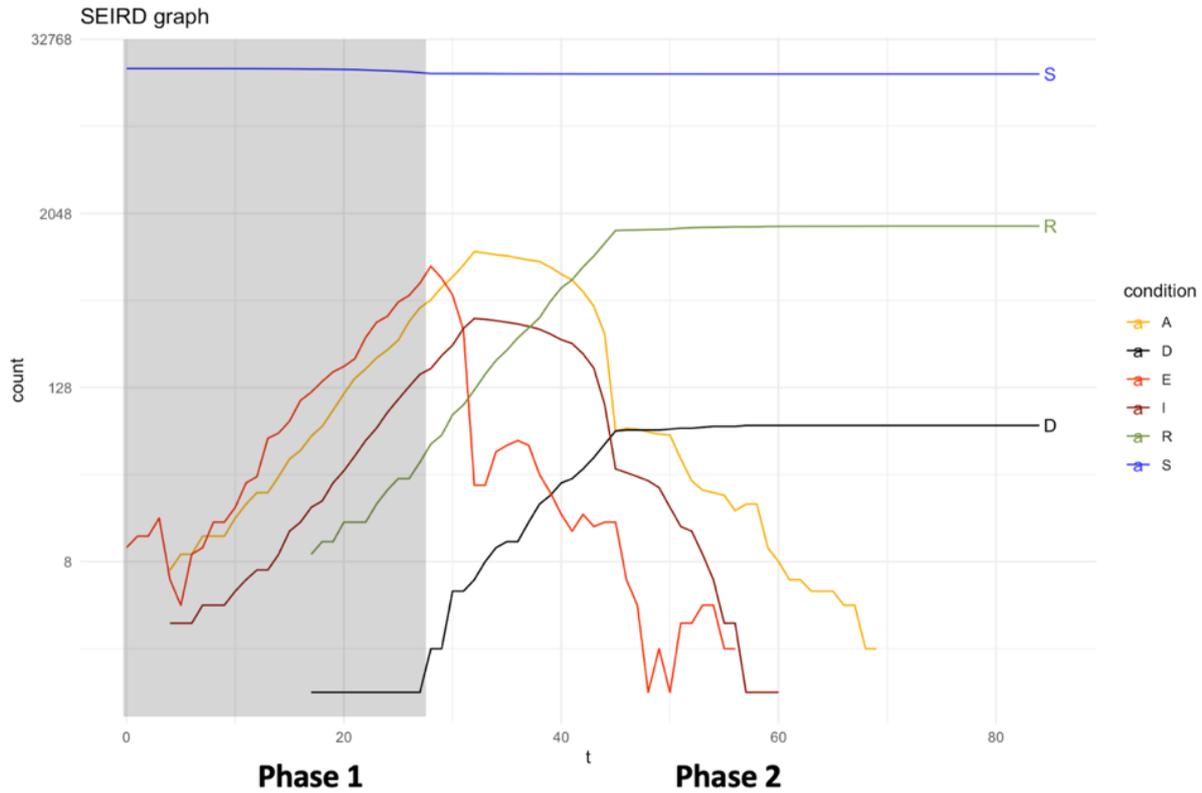

## 9.2 Simulation results

We present the main results obtained in the simulation exercise. Using the artificial population generated as described in the previous section, we considered the situation of a repeated sampling survey realized at three moments in time, namely, at day 15 (during the beginning of Phase 1), at day 25 (still in Phase 1, but in a situation closer to a *plateau*) and at day 35, during the period of lockdown. The infection situations at the three timepoints are reported in Table 1, where the results of the samples in Groups A and B and their intersection (see Figure 3) are separated.

Table 1. True simulated population values of the infected people for the two groups and their intersection observed on different days.

| Groups | Day 15 | Day 25 | Day 35 |
|---|---|---|---|
| $Y_A$ | 42 | 374 | 1,041 |
| $Y_B$ | 126 | 875 | 1,432 |
| $Y_{AB}$ | 39 | 372 | 1,018 |
| Total infected | **129** | **877** | **1,455** |

For group A, we fixed the parameter $g = 0.9$, while for group B, the parameters $f$ and $v$ were fixed as follows: $f = 0; v = 12$.

The sample sizes obtained with such parameter definitions (both excluding and including the contacts) are reported in Table 2 by distinguishing between 4 sample situations, namely, (i) A1B2: when both the individuals belonging to Group A and all their contacts are sampled while in Group B, both the noninfected and all contacts are sampled; (ii) A1B3: when both the individuals belonging to Group A and their contacts are sampled, while in Group B, the noninfected are sampled with all contacts (with a maximum of $v = 12$); (iii) A2B2: when all individuals belonging to Group A but only a subset of their contacts are included in the sample, and in Group B, the noninfected are sampled with all their contacts; and, finally, (iv) A2B3: when all individuals belonging to Group A but only a subset of their contacts are included in the sample, while in Group B, the noninfected are sampled with all their contacts but only up to a maximum of $v = 12$ individuals. Note that on day 35, we have fewer contacts in the sample than on day 25 due to the lockdown measures considered.

**Table 2. Total number of sampling units with and without contacts on different days and in the various sampling schemes.**

| Day | Proportion of infected people in the population | Sampling units without contacts | Sampling scheme | Sampling units with contacts |
|---|---|---|---|---|
| 15 | 0.006 | 4,130 | A1B2 | 4,741 |
| | | | A1B3 | 4,736 |
| | | | A2B2 | 4,741 |
| | | | A2B3 | 4,736 |
| 25 | 0.042 | 4,198 | A1B2 | 7,650 |
| | | | A1B3 | 7,634 |
| | | | A2B2 | 7,650 |
| | | | A2B3 | 7,634 |
| 35 | 0.070 | 4,361 | A1B2 | 7,545 |
| | | | A1B3 | 7,514 |
| | | | A2B2 | 7,545 |
| | | | A2B3 | 7,514 |

*Sampling scheme description: A1 = All individuals in Group A and their contacts are totally sampled; A2 = All individuals in Group A and a subset of their contacts are sampled; B2 = A subset of Group B and all their contacts are sampled; B3 = A subset of Group B and all their contacts are sampled, but only up to a maximum of $v = 12$ individuals.*

The main results of the simulation are reported in Table 3, and they show that in all sampling settings, the relative bias of our scheme is very small, and our estimators dramatically outperform simple random sampling in terms of efficiency (the ratio of the standard error of the proposed estimator computed by the simulation to that of the HT estimator for simple random sampling without replacement). In particular, the relative bias is on the order of 0.01% during Phase 1, while during Phase 2, it depends on the adoption of a sampling scheme with high precision when both the individuals belonging to Group A and their contacts are included in the sampling process. In contrast, the relative bias obviously increases when only a subset of the contacts is observed. Furthermore, our method outperforms the simple random sample in terms of efficiency. Similar to the case of bias, the relative advantage of our scheme over the simple random sample with respect to efficiency is greatest in the case of the A1 sample scheme when all selected individuals and their contacts are included in the sample, while it is lowest in the case of A2 when only a subset of them is observed. Moreover, Table 3 also displays a decrease in the relative advantage of our method for the day 35 wave, where due to the lockdown restrictions, the number of contacts is very limited.

The results presented here depend greatly on the particular settings of the (many) parameters involved in the simulation that describe different epidemic evolutions. To mitigate such subjectivity, we also run many other Monte Carlo experiments using different parameter values. Although they are available upon request, these results are not reported here for the sake of succinctness. However, they all confirm the same features: our method has a very low relative absolute bias and it is superior to the simple random sampling scheme in terms of efficiency.

Table 3: Results of the simulation study for the various sampling schemes on different days.

| Days (1) | Percentage of infected people in the population (2) | True population value (see Table 1) (3) | Sampling scheme (4) | Estimated total number of infected people (average over 500 simulations) (5) | $\alpha^*$ (6) | Standard error (7) | Coefficient of variation (%) $\frac{(7)}{(5)}x100$ (8) | Relative absolute bias $\frac{|(3)-(5)|}{(3)}$ (9) | Relative efficiency compared with that of the simple random sample without contacts (10) | Relative efficiency compared with that of the simple random sample with contacts (11) [a] |
|---|---|---|---|---|---|---|---|---|---|---|
| 15 | 0.0058 | 129 | A1B2 | 128.99 | 0.00 | 0.01 | 0.09 | 0.0001 | 0.0006 | 0.0006 |
|  |  |  | A1B3 | 128.99 | 0.00 | 0.01 | 0.09 | 0.0001 | 0.0006 | 0.0006 |
|  |  |  | A2B2 | 128.75 | 0.00 | 1.56 | 0.97 | 0.0019 | 0.0659 | 0.0718 |
|  |  |  | A2B3 | 128.75 | 0.00 | 1.56 | 0.97 | 0.0019 | 0.0659 | 0.0718 |
| 25 | 0.0394 | 877 | A1B2 | 876.83 | 0.00 | 0.17 | 0.05 | 0.0001 | 0.0027 | 0.0041 |
|  |  |  | A1B3 | 876.84 | 0.00 | 0.16 | 0.05 | 0.0001 | 0.0027 | 0.0040 |
|  |  |  | A2B2 | 876.90 | 1.00 | 0.48 | 0.08 | 0.0001 | 0.0080 | 0.0120 |
|  |  |  | A2B3 | 877.02 | 0.48 | 2.43 | 0.18 | 0.0000 | 0.0403 | 0.0605 |
| 35 | 0.0654 | 1.455 | A1B2 | 1,455.00 | 0.00 | 0.00 | 0.00 | 0.0000 | 0.0000 | 0.0000 |
|  |  |  | A1B3 | 1,455.00 | 0.00 | 0.00 | 0.00 | 0.0000 | 0.0000 | 0.0000 |
|  |  |  | A2B2 | 1,461.59 | 0.00 | 9.78 | 0.21 | 0.0045 | 0.1310 | 0.1895 |
|  |  |  | A2B3 | 1,461.59 | 0.00 | 9.78 | 0.21 | 0.0045 | 0.1310 | 0.1895 |

*Sampling scheme description: A1 = All individuals in Group A and their contacts are totally sampled; A2 = All individuals in Group A and a subset of their contacts are sampled; B2 = A subset of Group B and all their contacts are sampled; B3 = A subset of Group B and all their contacts are sampled, but only up to a maximum of $v = 12$ individuals.*

*Columns (10) and (11): The relative efficiency is computed as the ratio of the standard error of the proposed estimator (computed by the simulation) to that of the HT estimator for simple random sampling without replacement.*

## 10. Conclusions and future challenges

The aim of this paper is to draw the attention of researchers and decision makers to the need to observe the characteristics of the COVID-19 pandemic through a formal sampling design, thus overcoming the limitations of data collected on a convenience basis. Only in this way will we be able to produce both reliable estimates of the current situation and forecasts of the future evolution processes of epidemics so that we can make empirically grounded decisions about public health monitoring and surveillance, especially in the transition phase between the decline from the epidemic peak and the relaxation of quarantine measures.

In such a situation, it is essential to set up a system of data collection that allows for statistically valid comparisons over time and across different geographic areas by taking different economic, demographic, social, environmental and cultural contexts into account.

We believe that clear knowledge of the phenomenon is also necessary for the population to become aware of it and to adopt responsible behaviours. Trust and sharing must be grounded on a solid information base.

In comparison with other possible observational strategies, the proposal in this paper has three elements of strength:

- Relevance. The proposed sampling scheme, designed to capture most of the infected people through an indirect sampling mechanism, not only aims at providing a snapshot of the phenomenon at a single moment in time but is designed as a continuous survey that repeated in several waves over time. It also takes the different target variables in different stages of epidemic development into account and contributes to the implementation of a statistical surveillance system for the epidemic that could be integrated with existing systems managed by the health authorities.
- Accuracy (Eurostat, 2017). In this paper, the properties of the estimators have been formally proven and confirmed by analysing the results of a set of Monte Carlo experiments. The results guarantee the reliability of the estimators in terms of unbiasedness, and their efficiency is higher than that of a simple random sample.
- Timeliness. The sampling design is operable immediately, as this is required by the emergency we are experiencing. Indeed, this paper represents the statistical formalization of a recent proposal (Alleva et al., 2020) and has been accompanied by a technical note that describes the different phases into which it is divided, the subjects involved and the crucial aspects required for its success (Ascani, 2020).

Although our effort with regard to the pandemic has progressed during this phase of the emergency, there is room for much methodological and statistical research in terms of setting up statistical instruments for producing reliable and timely estimates of the phenomenon. Indeed, from a methodological point of view, while in this paper we have fully derived the properties of the estimators in the cross-sectional case, the properties in subsequent waves still need to be proven formally. Among other aspects to be developed, we mention those related to time and spatial correlations, which are useful both for modelling the phenomenon and for designing an efficient spatial sampling technique to achieve the same level of precision as that of the current method but with fewer sample units (Arbia and Lafratta, 2002). A specific extension of the spatial sampling techniques to be further developed is the use of capture/recapture techniques (Borchers, 2009; Lavallée and Rivest, 2012), which would require an overlap of the samples in Groups *A* and *B*. A further improvement to be explored could be derived from applying the Dorfmann procedure (Dorfmann, 1943) to reduce the number of tests and the cost of our method.

In addition to the methodological advances, other general aspects to be developed in concert with different specialists are the integration of the statistical system proposed here with the health authority's surveillance system for infected people and the use of their contact-tracking devices for statistical purposes. These devices could be useful both for the identification of contacts and for measuring the propensity of people to travel and the connected risks of doing so. To this end, it could be interesting to study the possibility of considering, within our framework, the proposal developed by Saunders-Hastings et al.

(2017), who addressed the problem of monitoring during a pandemic via a model approach. The need to monitor pandemics over time should represent the motivation for building an integrated surveillance system. This system should merge three different pieces of information within a unified database: (i) the information collected by the administrative institutions when receiving and treating individuals who have turned to the healthcare system; (ii) the statistical information collected on purpose with the aim of accurately measuring the diffusion of an infection; and finally, (iii) the data obtained through new sources for tracking the movements of people and their contacts.

A third extension of our proposal concerns the operational point of view. Indeed, the sampling design described in detail in Section 5 should be accompanied by the definitions of some key points:

- A control room that ensures the necessary inter-institutional collaboration for guiding field operations (Health Authorities at the national and regional levels, Statistical Offices, others).
- An effective information campaign to promote participation among the population; the required legal framework to assure the collection and analysis of personal data.
- A medical testing procedure to consider for the selected population (swabs, blood testing and DNA mapping).
- The geographical-temporal estimation domains of interest and the sample dimensions on the basis of the information needs and the available financial and organizational resources.
- The frequency of sampling for Groups *A* and *B*, as well as the length of stay in the panel of group *B*.
- The sociodemographic characteristics, living conditions and mobility behaviours to be collected at the individual and family levels to shed light on relative risks and to evaluate the effects of the policies adopted for modifying the evolution of the epidemic.

This can only be achieved if epidemiologists, virologists, and administrators of healthcare institutions work in conjunction with experts in mathematical and statistical modelling and forecasting and experts in the evaluation of public policies.

We designed the sampling mechanism considering the Italian situation, and we proved its feasibility by defining the previous key points to estimate the times and costs of our method (Alleva et al., 2020)[12]. In adopting the suggested strategy, different countries may require adjustments to take the peculiarities of their specific health system and institutional framework into account. For this research direction, the contributions of the National Statistical Offices, as well as common actions and the sharing of experiences at the European and worldwide levels, will be essential.

---

[12] The sample size required to assure a certain level of accuracy for the estimates depends on the base rate of infection. The unit cost of administering the swab and serological tests relies on the level of involvement in the survey by the public health authorities. The total cost depends on the length and the periodicity of the panel survey. For Italy, we estimated the cost of data collection at the national and regional levels (21 regions), for a case with 3 months of monitoring, a panel survey every 15 days and a base rate of infection of 0.04. With regard to Groups A and B, the sample sizes are 1,000 and 1,200 units, and this implies requirements of 6,000 and 7,200 swabs and total costs of 210,000 and 252,000 euros, respectively.

**Acknowledgements.** We are very grateful to Mike Hidiroglou, Pierre Lavallée and Giovanna Ranalli for their challenging discussion, careful reading of our paper, and useful suggestions, all of which have helped us improve the quality of our proposal. We also acknowledge the comments and suggestions received from Francisco Lima and Pedro Campos, President and Director of the Methodology Department at the Portuguese National Statistical Office, and from João Lopes from the same Department.

# APPENDIX

For the estimator in Equation (39), the following model can be assumed:

$(A1)\ E_M(y_k) = \mu, V_M(y_k) = \mu(1-\mu)$ for $k = 1, \ldots, N$,

where $E_M$ and $V_M$ denote the expectation and variance of the model, respectively. According to the above model, the anticipated variance of the estimate $\hat{Y}_{HT,SRS}$ may be defined as (Falorsi and Righi, 1915, Appendix 4):

$$(A2)\ AV(\hat{Y}_{HT,SRS}) = \frac{N}{N-1}\sum_{k=1}^{N}\left(\frac{N-n}{n}\right)V_M(y_k) = \frac{N}{N-1}\sum_{k=1}^{N}\frac{N}{n}\left(\frac{N-n}{N}\right)\mu(1-\mu)$$

$$(A3)\qquad \cong \sum_{k=1}^{N}\frac{N}{n}\mu(1-\mu) = \sum_{k=1}^{N}\frac{1}{f}\mu(1-\mu) = \frac{N}{f}\mu(1-\mu).$$

The approximation in $(A3)$ holds for large $N$ and small $f$ values.

For the estimators in Equation (41), we can introduce the following model:

$(A4)\ E_M(y_k) = \vartheta, V_M(y_k) = \vartheta(1-\vartheta)$ for $k \in U_{yl}$.

Furthermore, for the variables $y_j\ (L_{Cj} \geq 1)$ and $y_j\ (L_{vj} \geq 1)$, we can adopt the hypothesis that the probabilities

(A5) $P(L_{Cj} \geq 1 | k \in U_v \cap l_{k,j} = 1) \cong \gamma_A$

$P(L_{vj} \geq 1 | k \in U_C \cap y_{k,j} \, l_{k,j} = 1) \cong \gamma_B$

are roughly constant. Then, we may derive the following models:

(A6) $E_M[y_j(L_{Cj} \geq 1)] = \vartheta \gamma_A, V_M[y_j(L_{Cj} \geq 1)] = \vartheta \gamma_A (1 - \vartheta \gamma_A)$ for $k \in U_v \cap l_{k,j} = 1$,

$E_M[y_j(L_{vj} \geq 1)] = \vartheta \gamma_B, V_M[y_j(L_{vj} \geq 1)] = \vartheta \gamma_B (1 - \vartheta \gamma_B)$ for $k \in U_C \cap y_{k,j} \, l_{k,j} = 1$.

Let us assume that the number of contacts $L$ is roughly constant in $U$. Then, the total number of contacts derived from the two frames ($U_v$ and $U_C$) are:

$$TL_v = \sum_{k=1}^{P_v N} \sum_{j=1}^{N} l_{k,j} = P_v N L, \quad TL_C = \sum_{k=1}^{(1-P_v)N} y_k \sum_{j=1}^{N} l_{k,j} \cong \mu(1 - P_v)NL.$$

By considering the reasonable assumptions of a uniform distribution for the contacts of all the units, we have:

$$L_{vj} \cong \frac{P_v NL}{P_v N} = L \text{ and } L_{Cj} \cong \frac{\mu(1 - P_v)NL}{(1 - P_v)N} = L\mu.$$

Adopting the model in (A1) for the case where we do not know if the unit $k \in U_{yl}$ or the models in (A4) for $k, j \in U_{yl}$, the estimates $\hat{Y}_{A,SRS}$ and $\hat{Y}_{B,SRS}$ can be approximated by:

(A7) $\hat{Y}_{A,SRS} \cong \sum_{k=1}^{P_v n} \frac{1}{f} \sum_{j=1:j \in U_k}^{L} \frac{1}{L} y_j, \quad \hat{Y}_{B,SRS} = \sum_{k=1}^{(1-P_v)n} \frac{1}{f} y_k \sum_{j=1:j \in U_k}^{L} \frac{1}{L\mu} l_{k,j} y_j,$

$\hat{Y}_{AB,SRS}^A = \sum_{k=1}^{P_v n} \frac{1}{f} \sum_{j=1:j \in U_k}^{L} \frac{1}{L} y_j \gamma_A,$

$\hat{Y}_{AB,SRS}^B = \sum_{k=1}^{(1-P_v)n} \frac{1}{f} y_k \sum_{j=1:j \in U_k}^{L} \frac{1}{L\mu} y_j \gamma_B.$

Then, for the estimates derived from $S_A$, we have:

(A8) $AV(\hat{Y}_{A,SRS}) \cong \sum_{k=1}^{P_v N} \frac{1}{f} \sum_{j=1:j \in U_k}^{L} \frac{1}{(L\vartheta)^2} V_M(y_j) = \sum_{k=1}^{P_v N} \frac{1}{f} \sum_{j=1:j \in U_k}^{L} \frac{1}{L^2} \vartheta(1 - \vartheta)$

$= \sum_{k=1}^{P_v N} \frac{1}{f} \frac{L}{(L\vartheta)^2} \vartheta(1 - \vartheta) = \frac{P_v N}{f} \frac{1}{L} \vartheta(1 - \vartheta),$

$AV(\hat{Y}_{AB,SRS}^A) = \sum_{k=1}^{P_v N} \frac{1}{f} \sum_{j=1:j \in U_k}^{L} \frac{1}{L^2} \gamma_A \vartheta(1 - \gamma_A \vartheta) = \frac{P_v N}{f} \frac{1}{L} \gamma_A \vartheta(1 - \gamma_A \vartheta)$

$ACov(\hat{Y}_{A,SRS}, \hat{Y}_{AB,SRS}^A) = \sum_{k=1}^{P_v N} \frac{1}{f} \frac{L}{L^2} Cov_M[y_j(L_{Cj} \geq 1), y_j] = \frac{P_v N}{f} \frac{1}{L} \gamma_A \vartheta(1 - \vartheta).$

Combining the above results, we have:

(A9) $AV(\hat{Y}_{A,SRS} - \alpha \hat{Y}_{AB,SRS}^A) = AV(\hat{Y}_{A,SRS}) + \alpha^2 AV(\hat{Y}_{AB,SRS}^A) - 2\alpha ACov(\hat{Y}_{A,SRS}, \alpha \hat{Y}_{AB,SRS}^A)$

$\cong \frac{P_v N}{f} \frac{1}{L} \vartheta(1 - \vartheta) + \alpha^2 \frac{P_v N}{f} \frac{1}{L} \gamma_A \vartheta(1 - \gamma_A \vartheta)$

$- 2\alpha \frac{P_v N}{f} \frac{1}{L} \gamma_A \vartheta(1 - \vartheta)$

$$= \frac{P_v N}{f} \frac{1}{L} \vartheta[(1-\vartheta)(1-2\alpha\gamma_A) + \alpha^2 \gamma_A(1-\gamma_A \vartheta)].$$

To evaluate the anticipated variances and covariances for the estimates derived from $S_B$, we have to preliminarily consider these results:

(A10) $E_M(y_k y_j) = \Pr(y_k = 1) E_M(y_j | y_k = 1) + [1 - \Pr(y_k = 0)]0 = \mu\vartheta$ for $k \in U_C \cap j \in U_k$,

$V_M(y_k y_j) = E_M\left[(y_k y_j)^2\right] - \left[E_M(y_k y_j)\right]^2 = \mu\vartheta(1 - \mu\vartheta).$ for $k \in U_C \cap j \in U_k$,

$V_M[y_k y_j (L_{vj} \geq 1)] = \mu\vartheta\gamma_B(1 - \gamma_B \mu\vartheta),$

$Cov_M[y_k y_j (L_{vj} \geq 1), y_k y_j] = \mu\vartheta\gamma_B(1 - \mu\vartheta).$

Then, we have:

(A11) $AV(\hat{Y}_{B,SRS}) \cong \sum_{k=1}^{(1-P_v)N} \frac{1}{f} \sum_{j=1: j \in U_k}^{L} \frac{1}{(L\mu)^2} V_M(y_k y_j) = (1-P_v)N \frac{1}{fL\mu} \vartheta(1-\mu\vartheta),$

$AV(\hat{Y}_{AB,SRS}^B) \cong (1-P_v)N \frac{1}{fL\mu} \vartheta\gamma_B(1-\gamma_B \mu\vartheta)$

$ACov(\hat{Y}_{B,SRS}, \hat{Y}_{AB,SRS}^B) \cong (1-P_v)N \frac{1}{fL\mu} \vartheta\gamma_B(1-\mu\vartheta).$

Combining the above results, we have:

(A12) $AV(\hat{Y}_{B,SRS} - (1-\alpha)\hat{Y}_{AB,SRS}^A)$

$= AV(\hat{Y}_{B,SRS}) + (1-\alpha)^2 AV(\hat{Y}_{AB,SRS}^B) - 2(1-\alpha)ACov(\hat{Y}_{B,SRS}, \alpha\hat{Y}_{AB,SRS}^B)$

$= (1-P_v)N \frac{1}{fL\mu} \vartheta[(1-\mu\vartheta) + (1-\alpha)^2 \gamma_B(1-\gamma_B \mu\vartheta) - 2(1-\alpha)\gamma_B(1-\mu\vartheta)].$